\documentstyle[12pt]{article}

\newcommand{\ie}{{\it i.e.}}
\newcommand{\eg}{{\it e.g.}}

\newcommand{\lsim}{\buildrel < \over {_\sim}}
\newcommand{\order}{{\cal O}}
\newcommand{\ccbar}{c\bar{c}}

\newcommand{\qqbar}{q\bar{q}}
\newcommand{\jp}{J/\psi}
\newcommand{\as}{$\alpha_s$}
\newcommand{\BRone}{{\rm Br}(\chi_1 \to \psi \gamma)}

\newcommand{\rhII}{{\varrho}_{11}}
\newcommand{\rhOO}{{\varrho}_{00}}

\newcommand{\RP}{|R'_P(0)|^2}
\newcommand{\als}{\alpha_s}

\newcommand{\PLB}[3]{\mbox{}Phys. Lett. {\bf B{#1}}, {#2} ({#3})}
\newcommand{\NPB}[3]{\mbox{}Nucl. Phys. {\bf B{#1}}, {#2} ({#3})}
\newcommand{\PRL}[3]{\mbox{}Phys. Rev. Lett. {\bf {#1}}, {#2} ({#3})}
\newcommand{\PRD}[3]{\mbox{}Phys. Rev. {\bf D{#1}}, {#2} ({#3})}
\newcommand{\ZPC}[3]{\mbox{}Z. Phys. {\bf C{#1}}, {#2} ({#3})}

\newcommand{\etal}{{\em et al.}}

\begin{document}

\thispagestyle{empty}
\begin{flushright}
   \vbox{\baselineskip 12.5pt plus 1pt minus 1pt
         SLAC-PUB-6576 \\
         July 1994 \\
         (T/E)
             }
\end{flushright}

\begin{center}
{\bf Polarization as a Probe to the Production Mechanisms
of Charmonium in $\pi N$ Collisions\footnote{Presented by W.-K. Tang
at QCD Tests Working Group, Workshop on the Future of High
Sensitivity Charm Experiments: Charm2000, Fermilab, Batavia, Il.,
June 7-9, 1994}}

\vskip 1\baselineskip

W.-K. Tang\footnote{Work supported by the Department
 of Energy, contract DE-AC03-76SF00515} and S. J. Brodsky$^2$\\
 {\normalsize \em
  Stanford Linear Accelerator Center} \\ {\normalsize \em Stanford
  University, Stanford, CA 94309}
\vskip 1\baselineskip
M. V\"anttinen and P. Hoyer \\
{\normalsize \em Department of
Physics} \\ {\normalsize  \em University of Helsinki, Finland}
\end{center}

\medskip

\begin{abstract}
\noindent
Measurements of the polarization of $\jp$ produced in pion-nucleus
collisions are in disagreement with
leading twist QCD prediction where  $\jp$
is observed to have negligible polarization whereas theory predicts
substantial polarization.
We argue that this discrepancy cannot be due to
poorly known
structure functions nor the  relative production
rates of $\jp$ and $\chi_J$.
The disagreement between theory and experiment suggests important
higher twist corrections, as has earlier been surmised
from the anomalous non-factorized nuclear
$A$-dependence of the $\jp$ cross section.
\end{abstract}

\section{Introduction}

One of the most sensitive tests of the QCD mechanisms for the
production of heavy quarkonium is the polarization of the $\jp$  in
hadron collisions. In fact, there are serious disagreements between
leading twist QCD
prediction \cite{BargerPRD31} and experimental data
\cite{Clark,Kourkoumelis,WA11,E673,Binon} on the production cross section
of `direct' $\jp$ and
$\chi_1.$ We would like to advocate that polarization of $\jp$ provides
strong constraints on the production mechanisms of $\jp$ and thus can
pinpoint the origin of these disagreements.

In this paper we will present some preliminary results on the
theoretical calculation of the polarization of $\jp$ in $\pi N$
collisions. The completed analysis will be published in a later
paper\cite{mhbt}. We found that the polarization of $\jp$ provides
important constraints on the nature of the production mechanisms and
urge that polarization measurement of $\jp$
should be included in the design of
future charm production experiment.

The paper is organized as follow. In section 2, we show that
from the experimental data on the production cross sections and leptonic decay
widths
of direct $\jp$ and $\psi'$, the long distance physics of formation of bound
states of $\ccbar$ can be separated from the short distance physics
of production of the $\ccbar$ pair. Thus, the perturbative analysis is
under control in calculating $\jp$ production even though the mass
of charm quark is not much larger than $\Lambda_{QCD}.$ Once the
validity of perturbative method is established, we
 calculate the production cross sections of
direct $\jp$, $\chi_1$ and $\chi_2$ in $\pi N$ collisions in PQCD.
 These
results are presented in section 3 and discrepancies are observed. We
show that, in comparison with the recent E705 and E672
data \cite{AntoniazziPRL70,Zieminski}, the predicted ratio of direct
$\jp$ production
compared to the
$\chi_2$ production is too low by a factor
of about 3. In addition the production ratio of production cross sections
of $\chi_1$ to $\chi_2$ is too low by a factor of 10 compared to data. A
similar
conclusion
 has been
reached in \cite{Schuler}. The polarization
data of $\jp$
\cite{Badier,Biino,Akerlof} allows us to make further conclusion of the
origin of the
disagreements.  In section 4, we find that even if the
relative production rates
of the $\jp,\  \chi_1$ and $\chi_2$ are adjusted
(using $K$-factors) to agree with the data,
the $\jp$ polarization data is still not reproduced. Therefore, the
discrepancies do not arise from an incorrect relative normalization of
the various channels and new production mechanisms are needed. We
will present our conclusion in the last section.

\section{Can direct $\jp$ production be calculated in PQCD?}
In leading twist QCD,
the production of the
$\jp$ at low transverse momentum occurs both `directly'
from the gluon fusion subprocess $gg \to \jp+g$ [Fig.~1a]
and indirectly
via the production and decay of $\chi_1$ and $\chi_2$ states.
These states
have
sizable decay branching fractions $\chi_{1,2} \to \jp+\gamma$
of 27\% and 13\%,
respectively.

\begin{figure}[htb]
\vspace{3.0in}
\caption {Fig.~1a shows direct $\jp$ production through $g g$ scattering. The
formation of
bound state is described by the wavefunction $\Psi^*_{\jp}(0)$ at the
origin. Fig.~1b shows
 leptonic decay of $\jp$ into $e^+e^-$ pair. The probability of
finding the $\ccbar$ pair is given by the wavefunction
$\Psi_{\jp}(0).$}
\label{factorization}
\end{figure}

In this model, we assume that
the non-perturbative physics, which is described by the wave function
at the origin in cases of production of $\jp$ and $\psi'$, is
separable from the perturbative hard subprocess, $\ie,$ factorization holds. As
the wave function at the
origin can be related to the leptonic decay amplitude [Fig.~1b], the ratio of
$\psi'$ to direct $\jp$ production can be expressed in terms of the
ratio of their leptonic decay width. More precisely, taking into account
of the phase space factor,
\begin{equation}
\frac{\sigma(\psi')}{\sigma_{dir}(\jp)} \simeq \frac{\Gamma(\psi'\to e^+e^-)}
  {\Gamma(\jp\to e^+e^-)} \frac{M_{\jp}^3}{M_{\psi'}^3} \simeq 0.24 \pm 0.03
  \label{psiprime}
\end{equation}
where $\sigma_{dir}(\jp)$ is the cross section for direct production of the
$\jp$. The
ratio (\ref{psiprime}) should hold for all beams and targets, independently of
the size of the higher twist corrections in producing the point-like $\ccbar$
state. The energy should be large enough for the bound state to form outside
the
target. The available data is indeed compatible with (\ref{psiprime}). In
particular, the E705 value \cite{AntoniazziPRL70} is $0.24$.
In Table 1, the ratio of $\psi'$ to direct $\jp$ production with different
projectiles is presented. They are all consistent with the value $0.24.$

\begin{table}[htb]
\begin{center}
\begin{tabular}{|c|c|c|c|}
  \hline
  & $\sigma(\psi')$ [nb] & $\sigma_{dir}(\jp)$
    & $\sigma(\psi')/\sigma_{dir}(\jp)$ \\ \hline
  $\pi^+$ & $22\pm 5$ & $97\pm 14$  & $0.23\pm 0.07$ \\ \hline
  $\pi^-$ & $25\pm 4$ & $102\pm 14$ & $0.25\pm 0.05$ \\ \hline
  $p$     & $20\pm 3$ & $89\pm 12$  & $0.23\pm 0.05$ \\ \hline
\end{tabular}
\end{center}
\caption{Production cross sections for $\psi'$,  direct
$\jp$ and their ratio in $\pi^+ N$, $\pi^- N$ and $p N$ collisions.
The data is from Ref.
\protect\cite{AntoniazziPRL70}.}
\end{table}

The anomalous nuclear target $A$-dependence observed for the $\jp$
is also seen for the $\psi'$ \cite{AldeJP}, so that the ratio (\ref{psiprime})
is indeed independent of $A$. Therefore, at high energies, the
quarkonium bound state forms long after the production of the $\ccbar$ pair
and the formation process is well described by the non-relativistic
wavefunction at the origin.

\section{Production rates of $\psi$ and $\chi_J$ states at leading twist}

In leading twist and to
leading order in \as,  $\jp$  production can be
computed from the convolution of hard subprocess cross section
$gg\rightarrow \jp g$, $gg\rightarrow \chi_{j}$, {\em etc.},
with the parton
distribution functions in the beam and target. Higher order corrections in \as,
and relativistic
corrections to the charmonium bound states, are unlikely to change our
qualitative conclusions at moderate $x_{F}$. Contributions from direct $\jp$
production, as well as
from indirect production via $\chi_1$ and $\chi_2$ decays, will be included.
Due
to
the small branching fraction $\chi_0 \to \jp+\gamma$ of 0.7\%, the contribution
from $\chi_0$ to $\jp$ production is expected (and observed) to be negligible.
Decays from the radially excited $2^3S_1$  state, $\psi' \to \jp + X$,
contribute
to  the total $\jp$ rate at the few per cent level and will be ignored here.

The $\pi N \to \chi_2+X$ production cross section to lowest order is
\begin{equation}
  \sigma(\pi N \to \chi_2+X;\ x_F>0) = \int_{\sqrt{\tau}}^1 \frac{dx_1}{x_1}
  F_{g/\pi}(x_1) F_{g/N}(\tau/x_1) \sigma_0(gg\to \chi_2) \label{sigchi2}
\end{equation}
where
$\tau=M_{\chi_2}^2/s$ and the quantity
$\sigma_0(gg\to \chi_2) = 16 \pi^2 \als^2 \RP / M_{\chi_2}^7$
\cite{BaierZPC19}. We restrict the $\chi_2$ momentum range to the forward CM
hemisphere $(x_F>0)$ in accordance with the available data, and use the
structure functions of Ref. \cite{OwensPRD30,OwensPLB266} evaluated at
$Q^2=M_{\chi_2}^2$. We also take the renormalization scale to be
$Q^2=M_{\chi_2}^2$.

The direct $\pi N \to \jp+X$ cross section is similarly given by
\begin{eqnarray}
  \sigma(\pi N \to \jp+X;\ x_F>0) & = & \int_\tau^1 dx_1 \int_{\tau/x_1}^1 dx_2
  \int_{\hat t_{\rm min}}^0 d\hat t F_{g/\pi}(x_1) F_{g/N}(x_2)
  \nonumber \\
  & & \times \frac{d\sigma}{d\hat t}(gg\to \jp+g) \label{sigjp}
\end{eqnarray}
where $\hat t$ is the invariant momentum transfer in the subprocess, and
\begin{equation}
  \hat t_{\rm min} = {\rm max} \left(
  \frac{x_2 M_{\jp}^2 - x_1 \hat s}{x_1 + x_2}, M_{\jp}^2 - \hat s \right).
\end{equation}
Eq. (\ref{sigjp}) also applies to the $\pi N \to \chi_1+X$ reaction, in which
case a sum over the relevant subprocesses $gg \to \chi_1 g$, $gq \to \chi_1
q$, $g\bar{q} \to \chi_1 \bar{q}$
and $\qqbar \to \chi_1 g$ is necessary. The differential cross sections
$d\sigma/d\hat t$ for all subprocesses are given in
\cite{BaierZPC19,GastmansWu}.

In Table 2 we compare the $\chi_2$ production
cross section, and the relative rates of direct $\jp$ and $\chi_1$ production,
with the data of E705 and WA11 on $\pi^-N$ collisions at $E_{lab}=300$ GeV
and 185 GeV \cite{AntoniazziPRL70}.

\begin{table}[htb]
\begin{center}
\begin{tabular}{|c|c|c|c|}
  \hline
  & $\sigma(\chi_2)$ [nb] & $\sigma_{dir}(\jp)/\sigma(\chi_2)$
    & $\sigma(\chi_1)/\sigma(\chi_2)$ \\ \hline
  Experiment & $188 \pm 30\pm 21$  & $0.54 \pm 0.11\pm 0.10$ & $0.70 \pm
  0.15\pm 0.12$
   \\ \hline
  Theory     & 72 & 0.19 & 0.069 \\ \hline
\end{tabular}
\end {center}
\caption{Production cross sections for $\chi_1$, $\chi_2$ and directly produced
$\jp$ in $\pi^- N$ collisions. The data from Ref.
\protect\cite{AntoniazziPRL70,E705}
include measurements at 185 and 300 GeV. The theoretical
calculation is at 300 GeV.}
\end{table}

The $\chi_2$ production rate in QCD agrees with the data within a
`$K$-factor' of order $2$ to $3$. This is within the theoretical uncertainties
arising from the $\jp$ and $\chi$ wavefunctions, higher order corrections,
structure functions, and the
renormalization scale. A similar factor is found between the lowest-order
QCD calculation and the data on lepton pair production \cite{BadierDY,Conway}.
On the
other hand, Table 2 shows a considerable discrepancy between the calculated and
measured relative production rates of direct $\jp$ and $\chi_1$, compared to
$\chi_2$ production. A {\em priori} we would expect the $K$-factors to be
roughly
similar for all three processes.
We conclude that leading twist QCD appears to be in conflict with the data on
direct $\jp$ and $\chi_1$ production. Although in Table 2 we have only compared
our
calculation with the E705 and WA11
$\pi^- N$ data, this comparison is representative of
the overall situation (for a recent comprehensive review see \cite{Schuler}).

\section{Polarization of the $\jp$}

The polarization of the $\jp$ is determined by the angular distribution of its
decay muons in the $\jp$ rest frame. By rotational symmetry and parity, the
angular distribution of massless muons, integrated over the azimuthal angle,
has
the form
\begin{equation} \frac{d\sigma}{d\cos\theta} \propto 1 + \lambda \cos^2
\theta \label{lambda}
\end{equation}
where we take $\theta$ to be the angle between the $\mu^+$ and the projectile
direction (\ie, we use the Gottfried--Jackson frame). The parameter $\lambda$
can
be calculated from the $\ccbar$ production amplitude and the electric dipole
approximation of radiative $\chi$ decays.

The electric dipole approximation of the radiative decay
$\chi_J \to \psi \gamma$ is exact in the heavy quark limit; \ie, when terms of
$\order(E_\gamma/m_c)$ are neglected. As a consequence, the heavy quark
spins are conserved in the decay, while the orbital angular momentum changes.

The  lowest order subprocess
    $g(\mu_1)g(\mu_2) \to \ccbar \to \chi_2(J_z)$
only produces  $\chi_2$ with $J_z=\pm 2$ states asumming that the transverse
momenta of the incoming
gluons are neglected. In the $J_z=\pm 2$ polarization state the spin and
orbital
angular momenta
of its constituent charm quarks are aligned, $S_z = L_z = \pm 1$. Since $S_z$
is
conserved in the radiative decay $\chi_2 \to \jp+\gamma$, it follows that
$J_z(\jp)=S_z= \pm 1$ ($L=0$ for the $\jp$). Thus the $\jp$'s produced
via $\chi_2$ decay are transversely polarized, \ie, $\lambda=1$ in
(\ref{lambda}). This result is exact if both the photon recoil and the
intrinsic
transverse momenta of the incoming partons are neglected. Smearing of the beam
parton's transverse momentum distribution by a Gaussian function
$\exp \left[ -(k_\perp/500 \; {\rm MeV})^2 \right] $ would bring
$\lambda$ down to $\lambda \simeq 0.85$.

{}From the $gg\to\jp+g$ amplitude we find for direct
$\jp$ production, $\pi N \to \jp+X \to \mu^+\mu^-+X$,
\begin{eqnarray}
  \frac{1}{B_{\mu\mu}}\frac{d\sigma}{dx_F d\cos\theta} =
  \frac{3}{64\pi}
  \int \frac{dx_1 dx_2}{(x_1 + x_2) s} F_{g/\pi}(x_1) F_{g/N}(x_2)
  \nonumber \\ \times
  \left[ \rhII + \rhOO + (\rhII - \rhOO) \cos^2 \theta \right]
  \label{direct_distribution}
\end{eqnarray}
where $B_{\mu\mu}$ is the $\jp\to\mu^+\mu^-$ branching fraction,
$x_F=2p^z_\psi/\sqrt{s}$ is the longitudinal-momentum fraction of the $\jp$,
and $\theta$ is the muon decay angle of Eq. (\ref{lambda}). The $\rhII, \rhOO$
are the density matrix
elements and can be found in  \cite{mhbt}.

For the $\pi N\to \chi_1+X \to \jp+\gamma+X \to \mu^+\mu^-+\gamma+X$ production
process we get similarly
\begin{eqnarray}
  \frac{1}{B_{\mu\mu}}\frac{d\sigma}{dx_F d\cos\theta}
  & = & \frac{3}{128\pi} \BRone
  \Sigma_{ij} \int \frac{dx_1 dx_2}{(x_1 + x_2) s} F_{i/\pi}(x_1) F_{j/N}(x_2)
  \nonumber \\ & & \times
  \left[ \rhOO^{ij} + 3\rhII^{ij} + (\rhOO^{ij} - \rhII^{ij}) \cos^2 \theta
  \right], \label{chi1_distribution}
\end{eqnarray}
where the density matrix elements for $ij$ = $gg$, $gq$ $g\bar{q}$ and $\qqbar$
scattering
are again given in \cite{mhbt}.

In Fig.~2a we show the predicted value of the parameter $\lambda$ of Eq.
(\ref{lambda})
in the GJ-frame as a function of $x_F$, separately for the direct $\jp$ and the
$\chi_{1,2} \to
\jp+\gamma$ processes. Direct $\jp$ production gives $\lambda \simeq 0.25$,
whereas the production via $\chi_1$ results in $\lambda \simeq -0.15$.

The $\lambda(x_F)$-distribution obtained when both the direct and indirect
$\jp$ production processes are taken into account is shown in Fig.~2b and
is
compared with the Chicago--Iowa--Princeton
 \cite{Biino}
and E537 data \cite{Akerlof}
 for 252 GeV $\pi W$
collisions and 150 GeV $\pi^- W$ collisions respectively.
 Our QCD calculation gives $\lambda \simeq 0.5$ for $x_F \lsim
0.6$, significantly different from the measured value $\lambda \simeq 0$.

\begin{figure}[htb]
\vspace{3.5 in}
\caption {CIP ($\bullet$) and E537 ($\circ$) data compared with theoretical
prediction.
Fig.~2a shows the parameter $\lambda$ from different contributions:
direct $\jp$, $\chi_{1,2}
\rightarrow \jp + \gamma$ processes. Solid curves shows the results with
the intrinsic transverse momentum
of the incoming partons neglected while the dashed curves have the beam
parton's transverse momentum modeled by a Guassian function
$\exp[-(k_{\perp}/500 \mbox{MeV})^2].$  Fig.~2b takes into account both the
direct and indirect $\jp$ production: without $K$ factors correction
(solid curve), and with $K$ factors correction (dashed curve).}
\label{polarizat}
\end{figure}

The discrepancies between the calculated and measured values of $\lambda$
is one further indication that the standard leading twist processes
considered here are not adequate for explaining charmonium production.
The $\jp$ polarization is particularly sensitive to the production
mechanisms and allows
 us to make further conclusions on
the origin of the disagreements, including the above discrepancies in
the relative production cross sections of $\jp$, $\chi_{1}$ and
$\chi_{2}$. If these discrepancies arise from an incorrect
relative normalization of the various subprocess contributions (\eg, due to
higher order effects), then we would expect the $\jp$ polarization to agree
with data when the relative rates of the subprocesses are adjusted according to
the measured cross sections of direct $\jp$, $\chi_1$ and $\chi_2$
production\footnote{In the case of Drell-Yan virtual photon production, it is
known that higher-order
corrections do not change the $\gamma^*$ polarization significantly
\cite{Chiappetta}, which makes it plausible to represent these corrections by a
simple multiplicative factor, which does not affect the polarization of the
photon.}. The dashed curve in Fig.~2b shows the effect of multiplying the
partial
$\jp$ cross sections with the required $K$-factors. The $\lambda$ parameter is
still predicted incorrectly over most of the $x_F$ range.

A similar conclusion is reached (within somewhat larger experimental errors) if
we compare our calculated value for the polarization of direct $\jp$
production,
shown in Fig.~2a,
with the measured value of $\lambda$ for $\psi'$ production. In analogy
to Eq.~(\ref{psiprime}), the $\psi'$ polarization data
should agree with the polarization of directly produced
$\jp$'s, regardless of the production mechanism.  Based on the angular
distribution of the muons from $\psi' \to \mu^+\mu^-$ decays in 253 GeV
$\pi^-W$
collisions, Ref. \cite{Heinrich} quotes $\lambda_{\psi'} = 0.02 \pm 0.14$ for
$x_F>0.25$, appreciably smaller than our QCD values for direct $\jp$'s in Fig.~
2a.

\section{Discussion}

We have seen that the $\jp$ and $\chi_1$ hadroproduction cross sections in
leading twist QCD are at considerable variance with the data, whereas the
$\chi_2$
cross section agrees with measurements within a reasonable $K$-factor of
$2$ to $3$. On
the other hand, the inclusive decays of the charmonium states based on  the
minimal
perturbative
final states ($gg$ and $\qqbar g$) have been studied in detail using
perturbation
theory \cite{Kwong,Kopke,Schuler}, and appear to work fairly well.
It is therefore improbable that the treatment of the $\ccbar$ binding
should require large corrections. This conclusion is supported by the fact that
the relative rate of $\psi'$ and direct $\jp$ production (Eq.~\ref{psiprime}),
which at high energies
should be independent of the production mechanism, is in agreement with
experiment.

In a leading twist description, an incorrect normalization of the
charmonium production cross sections can arise from large higher order
corrections or uncertainties in the parton distributions\cite{Schuler}.
Taking into account that the normalization may be wrong by as much as a factor
of 10 and that even such a $K-$factor does not explain the polarization
data of $\jp$,
a more likely
explanation may be that
there are important
higher-twist contributions to the production of the $\jp$ and $\chi_1$
as suggested in large $x_F$ case \cite{BHMT,HVS}.

Further theoretical work is needed to establish that the data on direct $\jp$
and $\chi_1$ production indeed can be described from  higher twist
mechanisms.
 Experimentally, it is important to check whether the
$\jp$'s produced indirectly via $\chi_2$ decay are transversely polarized. This
would show that $\chi_2$ production is dominantly leading twist, as we have
argued. Thus, the polarization of $\jp$ production from different
channels provides a very sensitive discriminant of different
production mechanisms.

\end{document}